\documentclass{rspublic}
\usepackage{amsmath}
\usepackage{amsfonts}
\usepackage{amssymb}
\usepackage[Symbol]{upgreek}
\usepackage[nointegrals]{wasysym}
\usepackage{mathrsfs}
\usepackage{graphicx}
\newcommand{\bfu}{{\bf u}}
\newcommand{\bfx}{{\bf x}}
\newcommand{\Upo}{{\Sigma}}
\newcommand{\bdy}{{\partial \Sigma}}
\begin{document}
\title[Planar Navier-Stokes Equations in Bounded Domain]{\large Viscous flow regimes in unit square: Part 3. Phenomena of leapfrogging and approaching among multiple vortex pairs}
\author[F. Lam]{F. Lam}
%
%\affiliation{}
%
\label{firstpage}
\maketitle
\begin{abstract}{Planar Navier-Stokes Equations; Vorticity; Stream Function; Non-linearity; Laminar Flow; Transition; Turbulence; Diffusion}

In the present note, we solved numerically the viscous vorticity equation of the initial-boundary value problem describing the classic Helmholtz phenomena of vortex interaction. In the leapfrogging of vortex pairs, we demonstrate the fact that there exists a variety of initial vortex configurations, such as the initial vortex core structures, the starting speeds, the lateral and longitudinal separations as well as the fluid viscosity. To simulate leapfrogging appears to be a straightforward task, based on the vorticity calculations for a number of initial vortices with different core structures. Indeed, the evaluation of the unsteady vortex interactions requires accurate numerical simulations, and the solutions are diverse and intriguing. In particular, the impact of two asymmetric approaching vortices can produce peeled-off fission whirls or cross-bred eddies of mushroom topology. Over the course of flow development, we are unable to define any unique consistent Reynolds number which may be used to classify the individual flows, because of the multiplicity of characteristic lengths and velocity scales. Our key observation is that the initial-boundary value problems of fluid motion do not necessarily imply the realisation of dynamically similar flows.  
\end{abstract}
%
%\tableofcontents
%
\section{Introduction}\label{intro}
The vorticity theory of potential flow was introduced by Helmholtz (1858) in the middle of the nineteenth century. At the end of his paper, he wrote (in Tait's translation): 

\begin{quote}
We can now see generally how two ring-formed vortex-filaments having the same axis would mutually affect each other, since each, in addition to its proper motion, has that of its elements of fluid as produced by the other. If they have the same direction of rotation, they travel in the same direction; the foremost widens and travels more slowly, the pursuer shrinks and travels faster, till finally, if their velocities are not too different, it overtakes the first and penetrates it. Then the same game goes on in the opposite order, so that the rings pass through each other alternately.

If they have equal radii and equal and opposite angular velocities, they will approach each other and widen one another; so that finally, when they are very near each other, their velocity of approach becomes smaller and smaller, and their rate of widening faster and faster. If they are perfectly symmetrical, the velocity of fluid elements midway between them parallel to the axis is zero. Here, then, we might imagine a rigid plane to be inserted, which would not disturb the motion, and so obtain the case of a vortex-ring which encounters a fixed plane.

\end{quote}

Leapfrogging of two coaxial circular ring vortices is a fluid motion in $3$ dimensions. By symmetry, the rings may be regarded as axi-symmetric so that, on any plane cut through the diameter of the rings, we observe multiple twin vortex pairs interacting with each other (see, for instance, Shariff {\it et al.} 1989; Riley \& Stevens 1993). If the diameter of the rings is large, and the local curvature the vortex line becomes so small that the vortices behave effectively as $2$-dimensional vortices. In laboratory, it is much easier and more practical to observe the leapfrogging on planes, see, for instance, Yamada \& Matsui (1978) or plate $79$ of van Dyke (1982). As every observation of the phenomenon must be in real fluid ($\nu>0$), this in turn suggests that the viscosity can never be negligible, no matter how small the viscosity may be. To produce identical and robust vortex rings in low viscosity flows is by no means a simple task, see Watanabe (1995). 

Thus numerical simulations of leapfrogging or vortex interaction in $2$ dimensions is relevant and may shed light on the complexity of the dynamics, with suppression of vorticity stretching. The unsteady viscous vorticity equations are solved for given leapfrog configurations of multiple vortices. This type of work has been extensively exploited in the past decades, see, for instance, the review by Cheng {\it et al.} (2015).

Likewise, the second phenomenon of vortex approaching has been relatively less studied; it is much harder to generate perfect symmetric rings, or to observe the subsequent interaction in {\it real fluids}. Particularly, high accurate discretisation schemes are required in numerical simulations, as the vortex motions have many degrees of symmetry, which may persist over time, and must be faithfully preserved. 
   
The objectives of the present paper are two-fold: to study the leapfrogging of a series of initial vortices of topology manifold in flows of small viscosity $\sim O(10^{-5})$; to examine Helmholtz's second phenomenon in more detail. We will demonstrate the fact that, it is almost impossible to identify a consistent Reynolds number for leapfrogging, due to the variety of characteristic lengths and velocity scales involved. With pre-specified initial vortex pairs, we show that the face-to-face impact of the vortices gives rise to a wealth of intricate dynamics, which relates to the basic building blocks in turbulence.    
\subsection*{Planar vorticity equation}
In 2 space dimensions, the equations of motion for incompressible flows are  
\begin{equation} \label{ns}
	\partial_t \bfu  + \nu \Delta \zeta = - (\bfu. \nabla ) \bfu  - \nabla p/\rho, \;\;\; \nabla.\bfu = 0.
\end{equation}
All symbols have their usual meanings in fluid dynamics. The no-slip boundary condition applies $\bfu(\bfx){=}0 \;{\mbox{for}}\; \bfx \in \bdy$, or on the walls of the unit square.
The dynamics asserts that the vorticity, $\zeta {=} \nabla {\times} \bfu$, must be a solenoidal quantity, or $\zeta$ is normal to the $x{-}y$ plane. Evoking the stream function $\psi$, we derive the governing equations of the vorticity-stream function formulation
\begin{equation} \label{vort}
\partial_t  \zeta - \nu \Delta \zeta  = -u \partial_x\zeta - v \partial_y \zeta =- \partial_y \psi \; \partial_x \zeta + \partial_x \psi \; \partial_y \zeta, \;\;\;  \Delta \psi = - \zeta,
\end{equation}
where, for $t\geq0$, the no-slip condition implies the homogeneous Dirichlet boundary condition $\psi_{\bdy}=0$. We seek the spatio-temporal evolution of the given initial data,
\begin{equation} \label{vt-ic}
	\zeta(\bfx,0)=\zeta_0(\bfx)= \nabla\times \bfu_0(\bfx) \;\;\; \bfx \in \Upo,
\end{equation}
where $\bfu_0$ is the initial velocity. Reciprocally, $\bfu_0$ may be recovered from $\Delta \psi_0=-\zeta_0$.
All the length, time and mass are normalised by the SI standards, so that the dynamic equations are in a dimensionless form. In practice, it is easier to assign the initial vorticity than to specify the solenoidal velocity satisfying the no-slip. The solutions of the dynamics are found by the numerical procedures as given in Lam (2018) while the pressure is determined by solving the following Poisson equation:
\begin{equation} \label{modp}
\Delta p = 2 \: \rho \: \big( u_x \: v_y - u_y\: v_x \big), 
\end{equation}
subject to Neumann boundary conditions, $\partial_x p$ on $x=0,1$, and $\partial_y p$ on $y=0,1$. These boundary data become available from (\ref{ns}) once the vorticity field is known.
\section{Leapfrogging} 
\subsection*{Doublet vortices in tandem}
\begin{figure}[ht] \centering
  {\includegraphics[keepaspectratio,height=4.5cm,width=12cm]{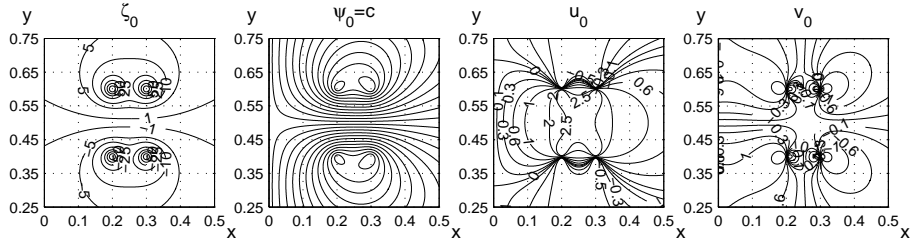}}
 \caption{Initial doublet vortices (\ref{doublet}). The wall effects are largely symmetric and weak compared to the self-induced motion. The choice of parameter $\kappa_D$ is to ensure incompressible flow field throughout the evolution. Note that $|\zeta_0|_{\max} \sim O(10^3)$ near the cores.  } \label{dbts} 
\end{figure}

\begin{figure}[ht] \centering
  {\includegraphics[keepaspectratio,height=11.5cm,width=11.5cm]{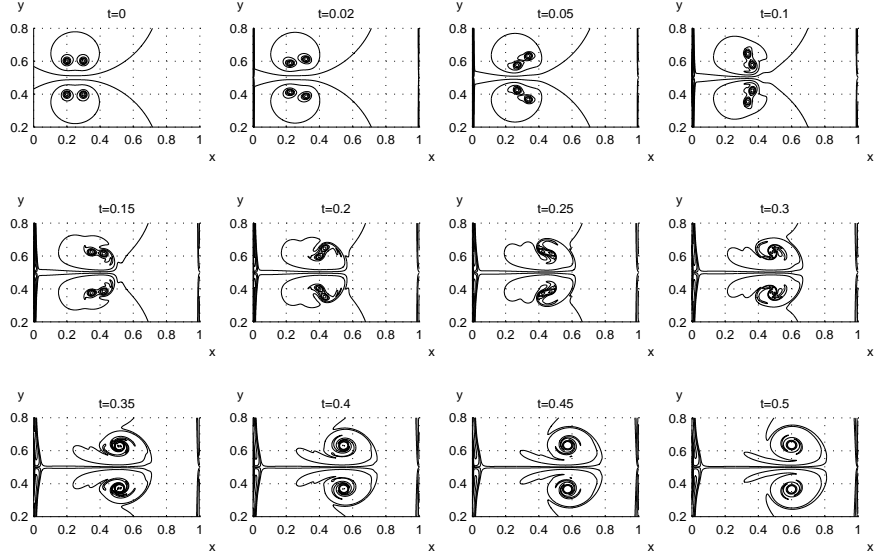}}
 \caption{Starting doublets in tandem and in staggered arrangement. Run data $\nu=10^{-4}$; grid $n=1024$; $\Delta t = 10^{-4}$. Displayed vorticity contours are $\pm250$, $\pm100, \pm75, \pm50, \pm25, \pm5$ and $\pm1$. The computations are largely straightforward but small zigzag ripples are found over $0.2 < t <0.3$ at low meshes ($n=256,512$), when the mixing appears to be intensive.  } \label{dbteq10k} 
\end{figure}

\begin{figure}[ht] \centering
  {\includegraphics[keepaspectratio,height=11.5cm,width=11.5cm]{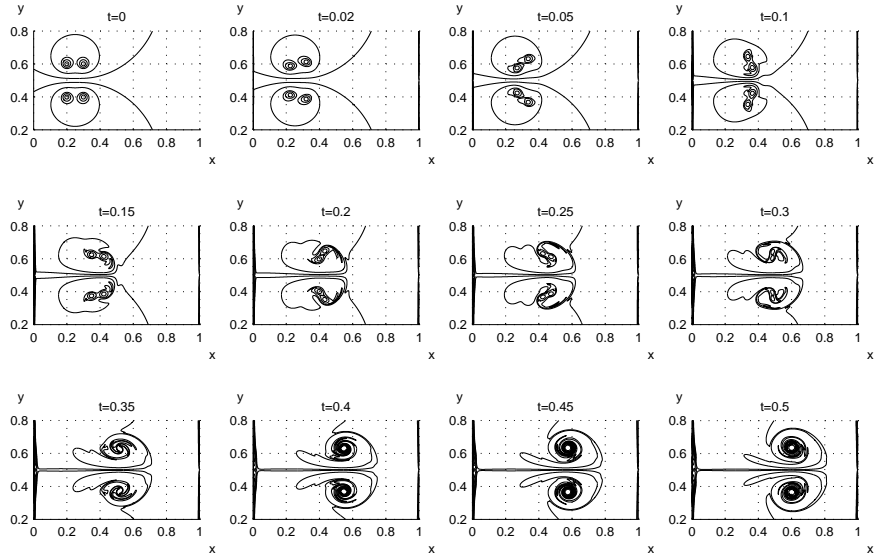}}
 \caption{Doublets at $\nu=2{\times}10^{-5}$; $n=2048$ (cf. figure \ref{dbteq10k}). The cores contain much stronger vorticity concentrations. The grid points ensure the smooth merging around $t=0.25$. The relative strengths of the two pairs must be strong enough for the leapfrogging to occur, given fixed separations between them. If the vorticity of the vortices is not intense and concentrated enough, diffusion dominates the motion so that the vortices would be simply fused and later annihilated by viscosity. This is a well-known experimental fact. Level contours are $\pm150$, $\pm75, \pm50, \pm25, \pm5$ and $\pm1$.} \label{dbteq50k} 
\end{figure}

The following expression describes isolated twin vortices or a doublet 
\begin{equation*} 
\eta_D(\alpha,\beta) = \frac{b}{\big(a+\alpha\big)^2+\big(b+\beta\big)^2/2+\kappa_D} + \frac{b}{\big(a+\alpha\big)^2+\big(b-\beta\big)^2/2+\kappa_D},
\end{equation*}
where $\kappa_D=10^{-4}$, $a=2x-1$, and $b=2y-1$.
Two pairs of such vortices are chosen as the initial data:
\begin{equation} \label{doublet}
	\zeta_0(\bfx)=\frac{1}{2}\big(\; \eta_D(0.6,0.2)\;+\;\eta_D(0.4,0.2)\;\big).
\end{equation}
Figure~\ref{dbts} shows the initial data (\ref{doublet}). The sense of the vorticity implies the pair will interact and proceed to the right. In figures \ref{dbteq10k} and \ref{dbteq50k}, the results of the present simulations are plotted.

In addition, the function $\eta_D$ gives us a leeway to define convenient data. The next expression,
\begin{equation} \label{dbtsm}
	\zeta_0(\bfx)=\frac{1}{2}\big(\; \eta_D(0.6,0.1)\;+\;\eta_D(0.4,0.2)\;\big),
\end{equation}
defines a staggered arrangement with an almost identical core structure (figure \ref{dbtsm10k}), while the following 
\begin{equation} \label{dbtlx}
	\zeta_0(\bfx)=\frac{1}{2}\big(\; \eta_D(0.6,0.1)\;+\;\eta_D(0.4,0.4)\;\big),
\end{equation}
specifies a set-up with larger separation of the right pair (figure \ref{dbtlx10k}).
\subsection*{Burgers vortices}
\begin{figure}[ht] \centering
  {\includegraphics[keepaspectratio,height=4.5cm,width=12cm]{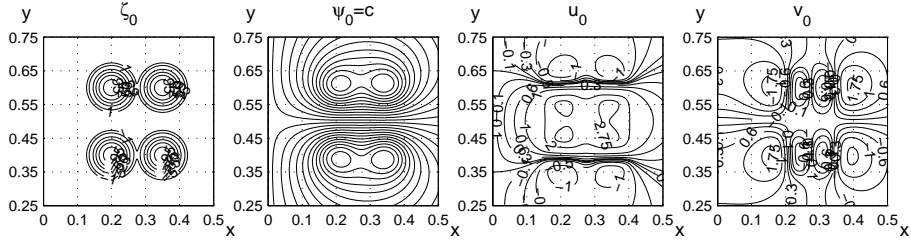}}
 \caption{Zoomed-in exponential vortex rings (\ref{burgers}).} \label{burgs} 
\end{figure}

\begin{figure}[ht] \centering
  {\includegraphics[keepaspectratio,height=11.5cm,width=11.5cm]{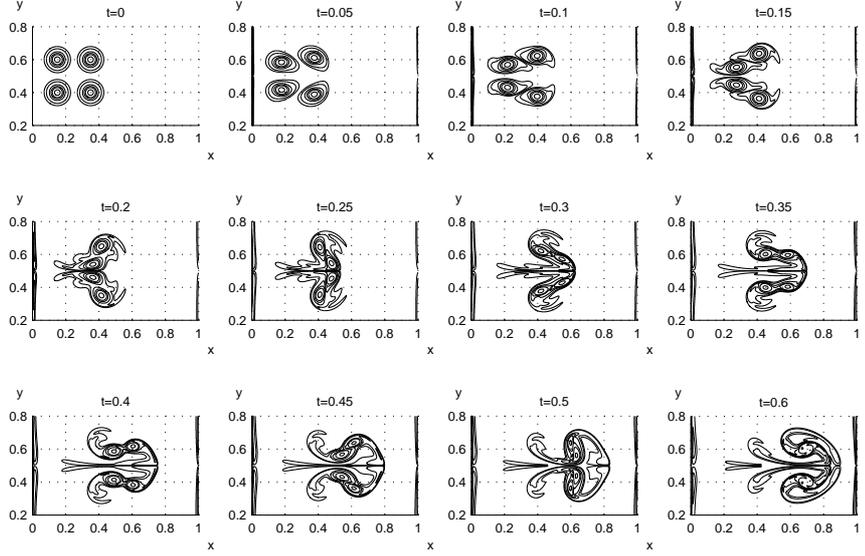}}
 \caption{Starting Burgers pairs (\ref{burgers}) at $\nu=10^{-4}$; grid $n=1024$; $\Delta t = 10^{-4}$. Displayed vorticity contours are $\pm125$, $\pm75, \pm50, \pm25, \pm5$ and $\pm1$. We notice that the core kernels, which dominate the evolution, show the lineaments unseen in those of the doublets.} \label{burgs10k} 
\end{figure}

\begin{figure}[ht] \centering
  {\includegraphics[keepaspectratio,height=11.5cm,width=11.5cm]{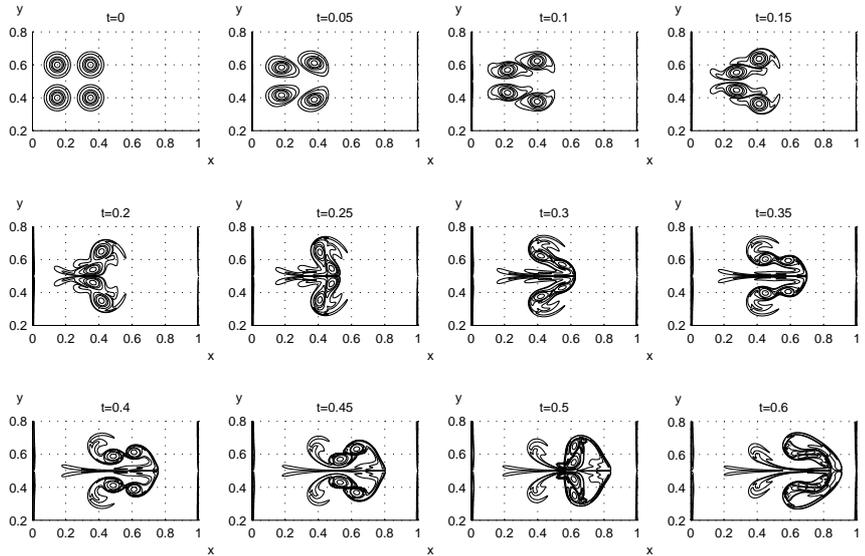}}
 \caption{Burgers pairs at a lower viscosity $\nu=2{\times}10^{-5}$; grid $n=2048$. With the identical contours of the previous plot, we see that the centre shears are more robust and are able to maintain their oval shape much longer (up to $t > 0.6$). } \label{burgs50k} 
\end{figure}

\begin{figure}[ht] \centering
  {\includegraphics[keepaspectratio,height=6cm,width=12cm]{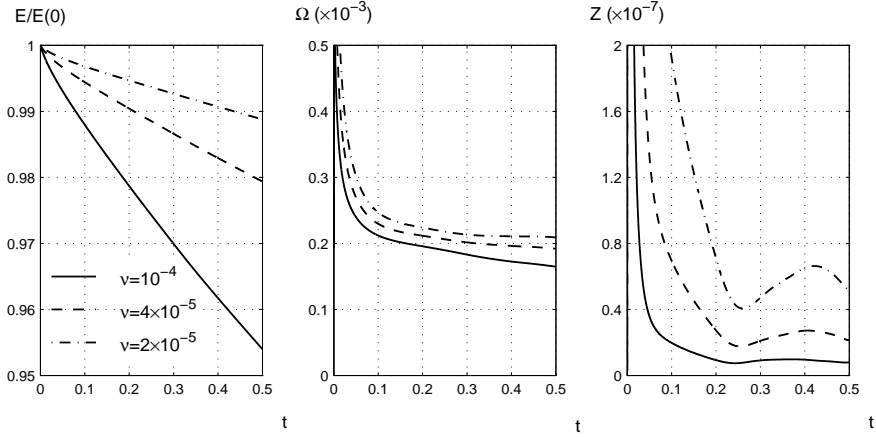}}
 \caption{Burgers vortices at three viscosity values. The $\nu$-dependence is obvious for the identical initial core geometry. In particular, the palinstrophy plot confirms our observation of the dissimilar amalgamations. } \label{burghist} 
\end{figure}

The next initial condition is a combination of the Burgers vortex, which is described by the expression
\begin{equation*}
	\eta_B(\alpha,\beta)=\frac{\pi}{4 l^2}\exp(-r_B^2/l^2),
\end{equation*}
where $r_B^2=(\alpha+2x-0.5)^2+(\beta+2y-1)^2$, and the core strength parameter, $l^2=0.005$, is fixed. We choose an initial set of two pairs of the vortices as follows:
\begin{equation} \label{burgers}
	\zeta_0(\bfx)=\eta_B(0.2,-0.2)-\eta_B(0.2,0.2)\;+\;\eta_B(-0.2,-0.2)-\eta_B(-0.2,0.2).
\end{equation}
The initial data are shown in figure~\ref{burgs}. Computational results are summarised in figures~\ref{burgs10k} to \ref{burghist}.
\subsection*{Lamb dipoles}
\begin{figure}[ht] \centering
  {\includegraphics[keepaspectratio,height=4.5cm,width=12cm]{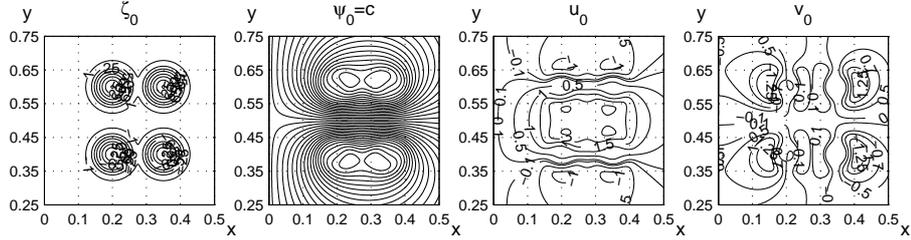}}
 \caption{Exponential rings (\ref{lamb}) have concentric bands of $|\zeta| \approx 65$ inside the cores.} \label{lamb0} 
\end{figure}

\begin{figure}[ht] \centering
  {\includegraphics[keepaspectratio,height=11.5cm,width=11.5cm]{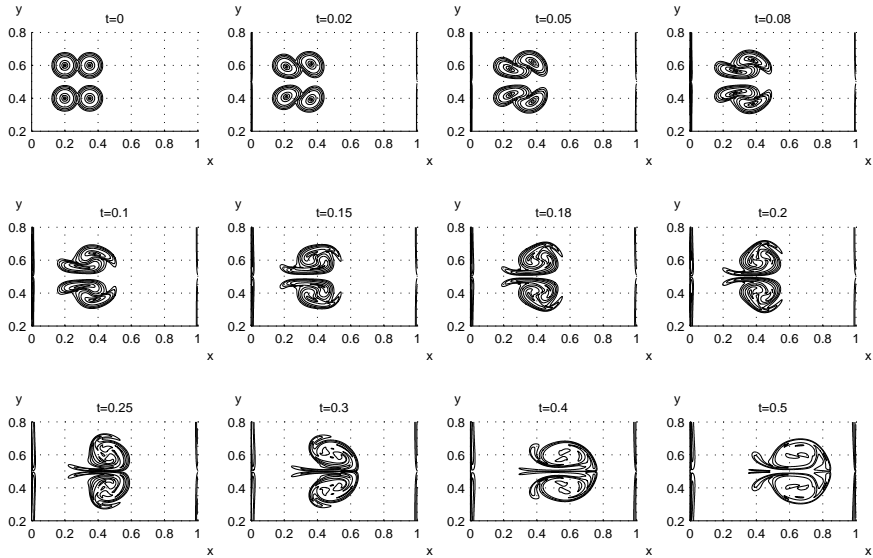}}
 \caption{Starting Lamb dipoles (\ref{lamb}) at $\nu=10^{-4}$; grid $n=1024$; $\Delta t = 10^{-4}$. Plotted iso-contours are $\pm100, \pm50, \pm20, \pm5$ and $\pm1$. Compared to the doublets and Burgers vortices, the marked differences lie at the core constructions where the initial shears are largely smeared-out at $t \sim 0.5$, resulting in a vortex pair with hollow cores.} \label{lamb10k} 
\end{figure}

\begin{figure}[ht] \centering
  {\includegraphics[keepaspectratio,height=11.5cm,width=11.5cm]{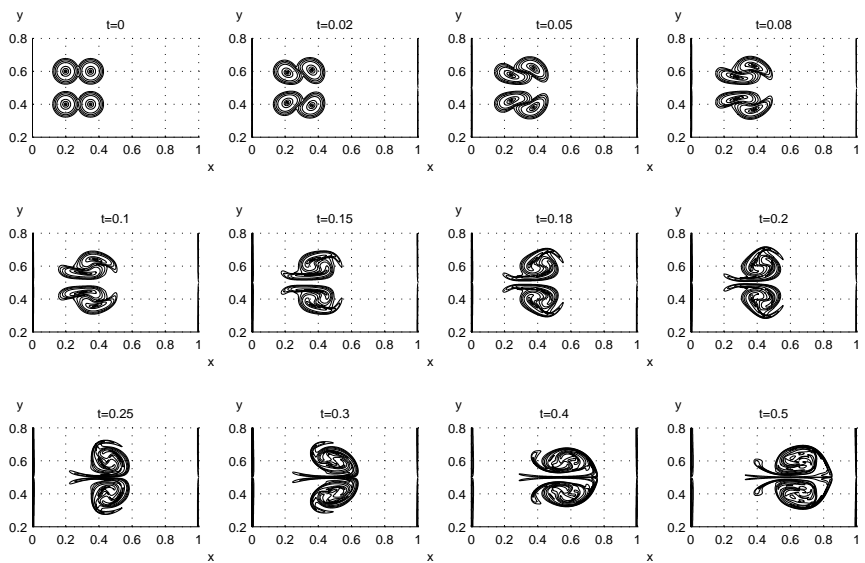}}
 \caption{Lamb dipoles at $\nu=2{\times}10^{-5}$; grid $n=2048$. Identical contours as in the preceding plot. Note that the induced viscous layers on the walls are extremely thin.  } \label{lamb50k} 
\end{figure}

\begin{figure}[ht] \centering
  {\includegraphics[keepaspectratio,height=6cm,width=12cm]{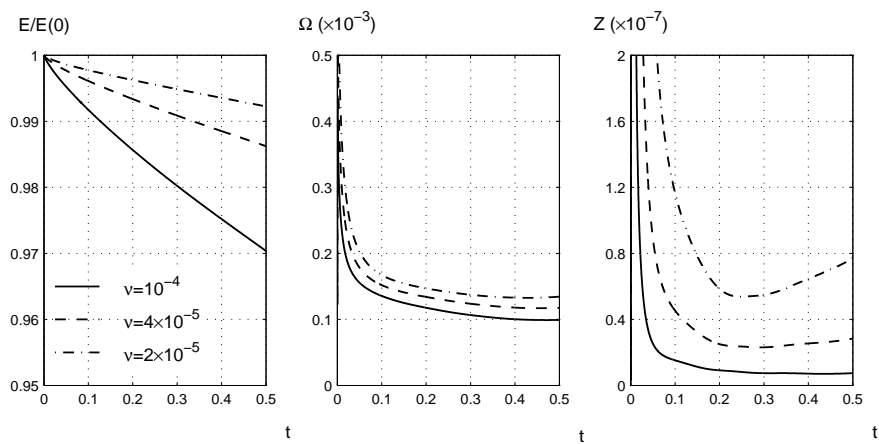}}
 \caption{Variation of Lamb dipoles as a strong function of viscosity.} \label{lambhist} 
\end{figure}

The starting eddies close to the left wall are given by
\begin{equation*} 
\sigma(\alpha,\beta) = \big(\;(2x-1+\alpha)^2+(2y-1+\beta)^2\;\big)/c^2,
\end{equation*}
where $c^2=0.01$. The parameters, $\alpha$ and $\beta$, are used to control the core location. An isolated vortex can be constructed:
\begin{equation*}
	\xi(\alpha,\beta)=(4\pi)^2 \:\sigma\exp\big(- \sigma^2 \big).
\end{equation*}
We seek the movement of two such pairs over time. The initial vorticity is expressed as the sum of the twins:
\begin{equation} \label{lamb}
\zeta_0(\bfx) =  \xi(0.6,0.2) - \xi(0.6,-0.2) \;+\;  \xi(0.3,0.2) - \xi(0.3,-0.2). 
\end{equation}
As indicated in figure~\ref{lamb0}, the present data have weaker and larger cores compared to the previous case, see figures~\ref{lamb10k} to \ref{lambhist}.
\section{Agglutination of vortices}
This is Helmholtz's second scenario: two identical or nearly identical rings with opposite sense of rotations travel to approach each other. Having met one another, the vortices appear to agglomerate, and the diameter increases due to mutual interaction. On any plane, cut normal to the rings, we observe the coalescence of point vortices while their separation grows.
\subsection*{Symmetric rings: approaching}
\begin{figure}[ht] \centering
  {\includegraphics[keepaspectratio,height=4.5cm,width=12cm]{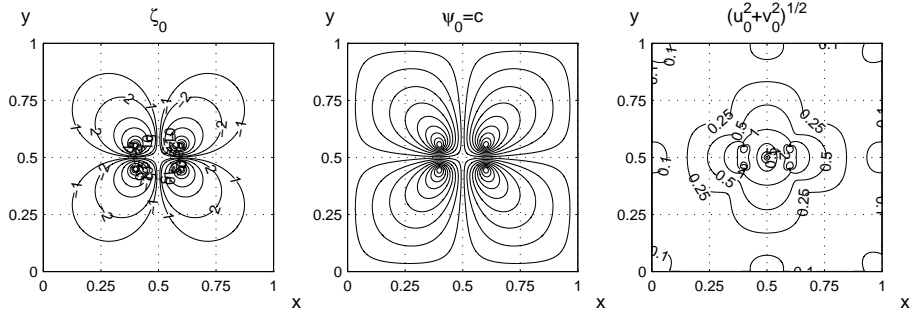}}
 \caption{Initial identical vortex rings (\ref{rings}). Under the mutual interaction of vorticity dynamics, the pair approach each other and rendezvous at the line $x=0.5$. The wall influence is clearly symmetrical and is expected to be minimal. } \label{hmtz} 
\end{figure}

The following simple function defines a $2d$ vortex ``ring'':
\begin{equation} \label{r1} 
	\eta_0(\alpha,\beta) = 
	\frac{y_0}{(x_0+\alpha)^2+(y_0-\beta)^2+\varepsilon_R}+\frac{y_0}{(x_0+\alpha)^2+(y_0+\beta)^2+\varepsilon_R},
\end{equation}
where $x_0=2x-1$, and $y_0=2y-1$, so as to fix the ring geometry into the unit square. The parameter, $\varepsilon_R=10^{-4}$, is used to specify the core structure of the vortices. We are interested in the development of two such rings:
\begin{equation} \label{rings} 
	\zeta_0(\bfx) = \eta_0(0.2,0.1)-\eta_0(-0.2,0.1).
\end{equation}
Figure~\ref{hmtz} illustrates the initial topology. Even in the limited space, we can still demonstrate what Helmholtz anticipated when two identical vortex systems collide with each other, at least over a limited time span. In laboratory, it is extremely hard to generate vortex rings of equal strength, and to observe the subsequent colliding sequence. On the other hand, it is challenging to numerically simulate the dynamics as the computations demand high accuracy.
\subsection*{Symmetric doublets}
In addition to the vortex ring geometry, we examine the head-on collision of two doublet pairs of equal strength initiated on line $y=1/2$:
\begin{equation} \label{edbt} 
	\zeta_0(\bfx) =  \frac{2 \pi y_f}{( x_f+1 )^2 + y_f^2 + \varepsilon_D}-\frac{2 \pi y_f}{ ( x_f-1 )^2 + y_f^2 + \varepsilon_D},
\end{equation}
where $\varepsilon_D=10^{-6}$ controls the centre singularity, $x_f=4x-2$, and $y_f=4y-2$.
\subsection*{Asymmetric doublets: peeling-off and merging}
\begin{figure}[ht] \centering
  {\includegraphics[keepaspectratio,height=4.5cm,width=12cm]{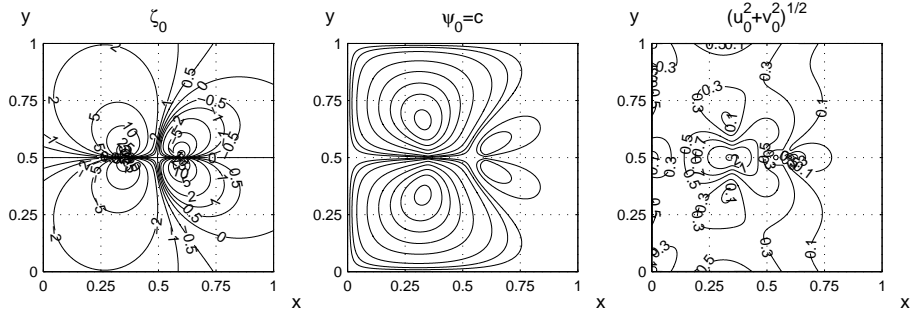}}
 \caption{Initial vortex doublet $\sigma_L:\sigma_R = 2:1$.} \label{adbt} 
\end{figure}

\begin{figure}[ht] \centering
  {\includegraphics[keepaspectratio,height=12.5cm,width=12.5cm]{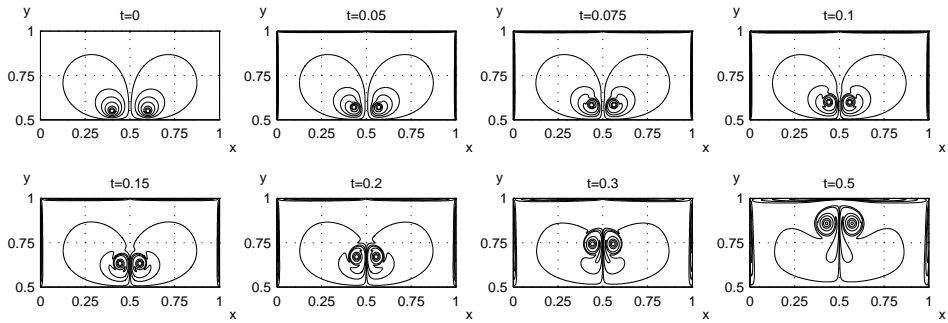}}
 \caption{Helmohltz's second scenario in real flows. Identical vortex rings (\ref{rings}) approach one another. Run at $\nu=10^{-4}$, $n=1024$. Even in this limited space, we manage to simulate what he anticipated in the last paragraph of his paper (i.e., the quote above). As the diameter of the coalesced vortices grows, the cores inflate and become diffused. Iso-vorticity contours are shown at $\pm200, \pm100$, $\pm50$, $\pm20, \pm10,\pm5 $ and $\pm1$. (Only the upper-half plots are presented in order to cut down the file size.) } \label{hdon10k} 
\end{figure}

\begin{figure}[ht] \centering
  {\includegraphics[keepaspectratio,height=12.5cm,width=12.5cm]{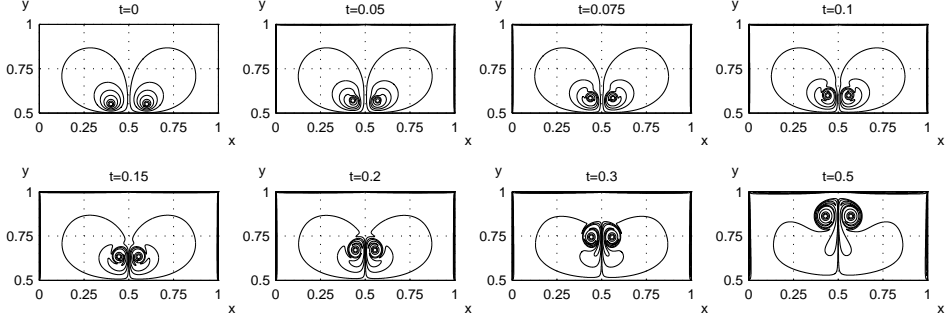}}
 \caption{As in figure~\ref{hdon10k}, run made at $\nu=2{\times}10^{-5}$ and $n=2048$. Of course, the collision-Treff of his version can only occur in ideal fluids. If we insert a solid plate along the line $x=1/2$ in our computations, boundary layers will form on either side of the plate, as the course of the vortices' motion has changed from horizontal to upward (and downward) after the coalescence (from $t \approx 0.05$ onward). Nevertheless, the cross-section of his agglutinated rings are of mushroom type. } \label{hdon50k} 
\end{figure}

\begin{figure}[ht] \centering
  {\includegraphics[keepaspectratio,height=12.5cm,width=12.5cm]{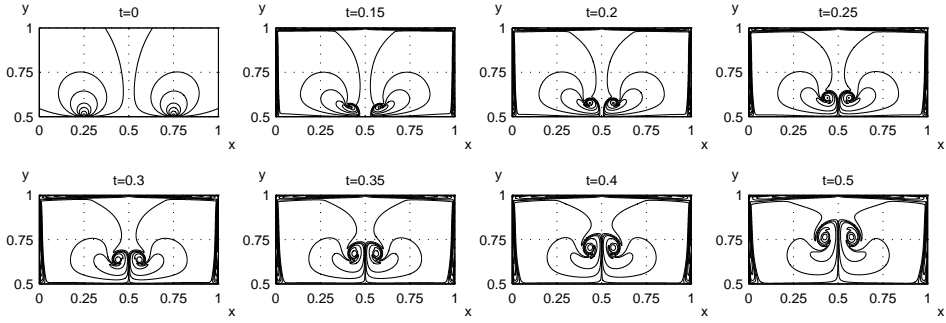}}
 \caption{Twin identical doublets (\ref{edbt}) approaching each other. Initial eddies (\ref{edbt}) located at $(0.25,0.5)$ and $(0.75,0.5)$. Computational runs at $\nu{=}10^{-4}$, $n{=}1024$, and $\Delta t {=}10^{-4}$. Displayed contours are $\pm100$, $\pm50$, $\pm30$, $\pm20, \pm10 $ and $\pm5$.} \label{aproch10k} 
\end{figure}

\begin{figure}[ht] \centering
  {\includegraphics[keepaspectratio,height=12.5cm,width=12.5cm]{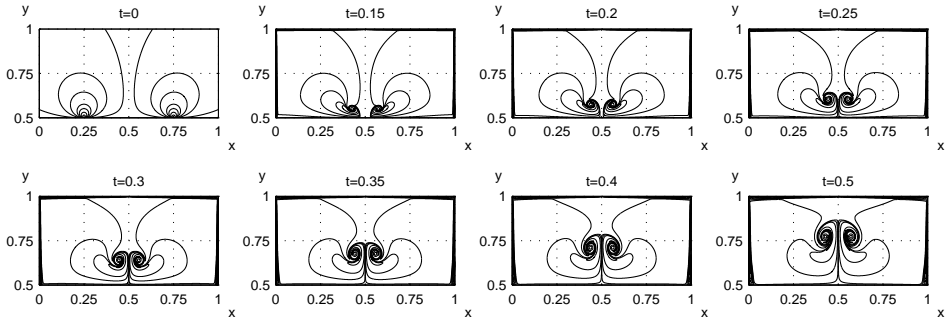}}
 \caption{Run at smaller viscosity $\nu=2{\times}10^{-5}$, $n=2048$. The mushrooms clearly have stronger cores compared to those in the preceding plot. (The wall layers are much thinner.) } \label{aproch50k} 
\end{figure}

\begin{figure}[ht] \centering
  {\includegraphics[keepaspectratio,height=12.5cm,width=12.5cm]{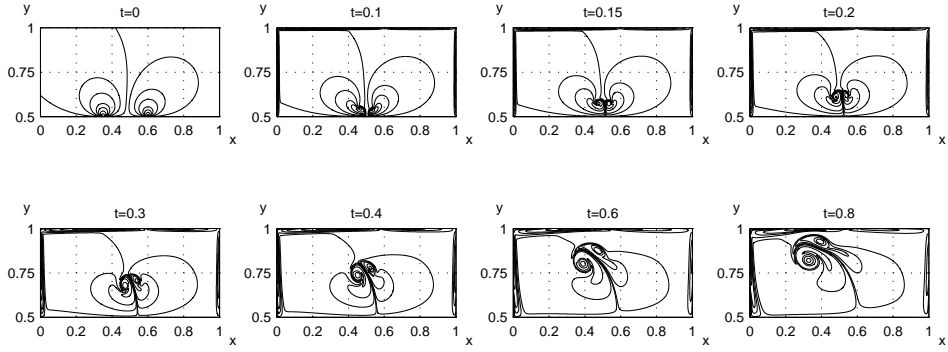}}
 \caption{After the head-on joust of asymmetric data (\ref{dbt}) with ratio $\sigma_L:\sigma_R = 2:3/2$. The branches of the vortices with opposite signs are self-blended to forge into cross-bred mushrooms, as expected. Computations carried out at $\nu=10^{-4}$, $n=768$. (Cf. the figures on page 24 of Samimy {\it et al.}, 2004.)} \label{hpear10k} 
\end{figure}

\begin{figure}[ht] \centering
  {\includegraphics[keepaspectratio,height=12.5cm,width=12.5cm]{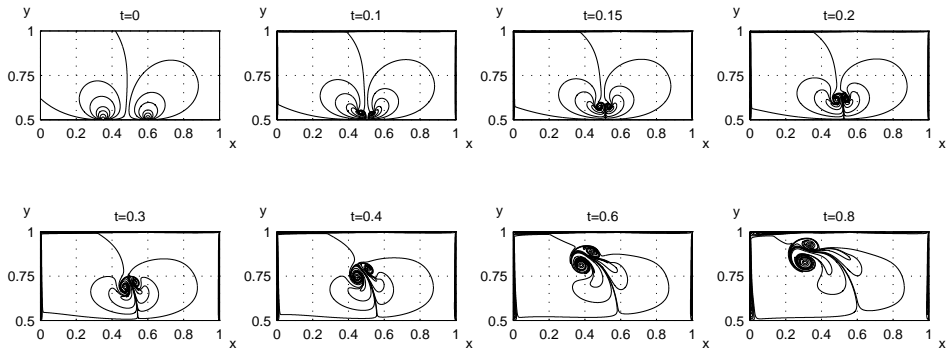}} 
 \caption{Upper-half snapshots at $\nu=10^{-5}$, $n=1536$, and $\Delta t = 10^{-4}$. Iso-contours are plotted at $\pm100$, $\pm50$, $\pm30$, $\pm20, \pm10$, $\pm5$ and $\pm1$. The impact time is similar to that of the preceding case, $0.1<t<0.15$. } \label{hpear100k} 
\end{figure}

\begin{figure}[ht] \centering
  {\includegraphics[keepaspectratio,height=12.25cm,width=12.25cm]{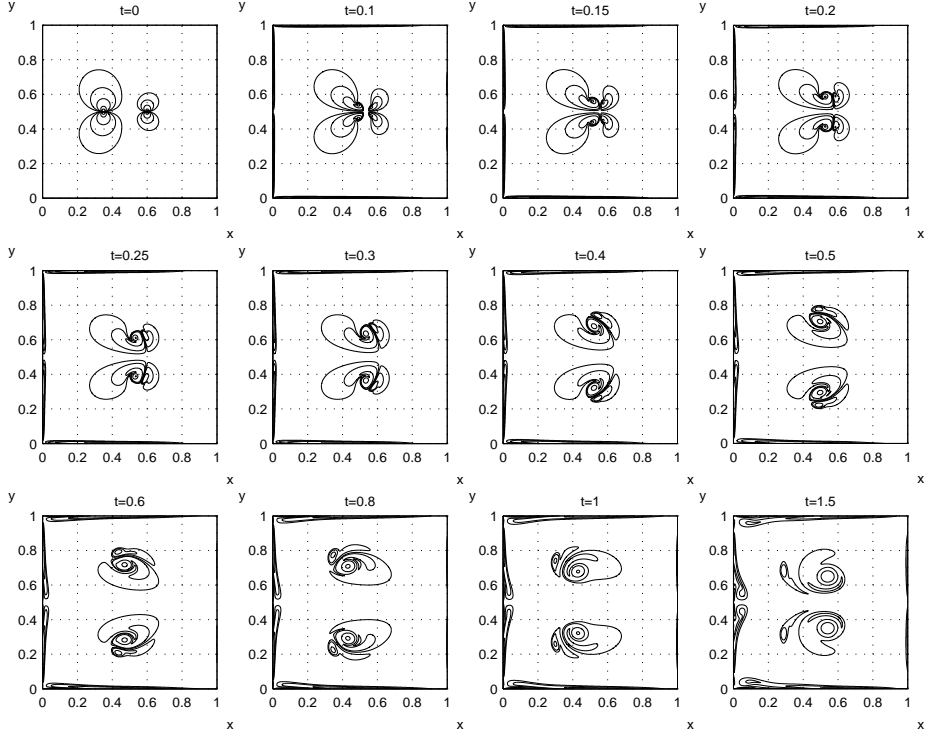}} 
 \caption{Dipoles peeling off (data \ref{dbt}) with weaker right pair $\sigma_L:\sigma_R = 2:1$, $\nu=10^{-4}$, $n=768$. Iso-contours are $\pm75$, $\pm50$, $\pm20, \pm10$, $\pm5$ and $\pm1$. For $t > 0.4$, the developed mushroom vortices are progressively tilted from $t{=}0.25$ and subsequently peeled-off by shearing. The final vorticity field appears to be populated by cleaved eddies. This is another mechanism of scale-multiplication. } \label{peff10k} 
\end{figure}

\begin{figure}[ht] \centering
  {\includegraphics[keepaspectratio,height=5.5cm,width=12cm]{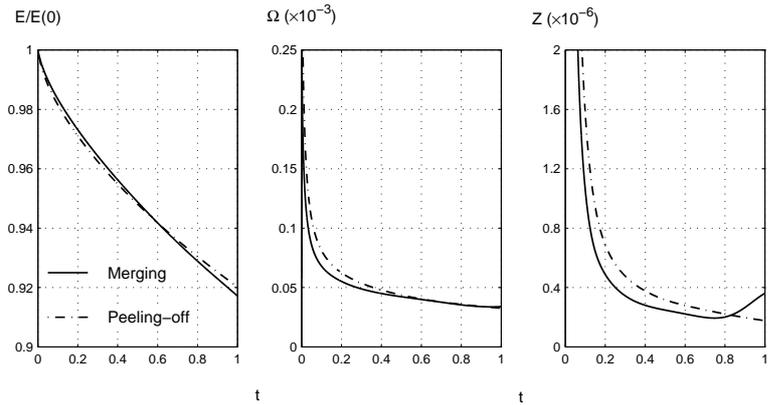}}
 \caption{Asymmetric doublets of figure~\ref{hpear10k} and figure~\ref{peff10k}. In this particular example, the integrated quantities do not reveal much of the detailed flow structures. However, they are crucial in the assessment of the mesh convergence so that essential flow data, such as the wall shears due to the vortex interaction, can be accurately predicted. } \label{aprhist} 
\end{figure}

A combination of the last data furnishes an example for vortices dynamics of variety:
\begin{equation} \label{dbt} 
	\zeta_0(\bfx) = \frac{\sigma_L\;\pi y_f}{( x_f+3/5 )^2 + y_f^2 + \varepsilon_D}-\frac{\sigma_R \; \pi y_f}{ ( x_f-2/5 )^2 + y_f^2 + \varepsilon_D}.
\end{equation}
Each of the two terms in (\ref{dbt}) is the familiar ``doublet singularity'' in the theory of potential flow in the limit of the parameter $\varepsilon_D \rightarrow 0$. We consider two cases for illustration: (1) $\sigma_L:\sigma_R {=} 2:3/2$; (2)$\sigma_L:\sigma_R {=} 2:1$. Figure~\ref{adbt} shows the second data. The calculations of configurations, (\ref{rings}), (\ref{edbt}) and (\ref{dbt}), are analysed and presented in figures~\ref{hdon10k} to \ref{aprhist}.
\section{Remarks}
\begin{figure}[ht] \centering
  {\includegraphics[keepaspectratio,height=6cm,width=12cm]{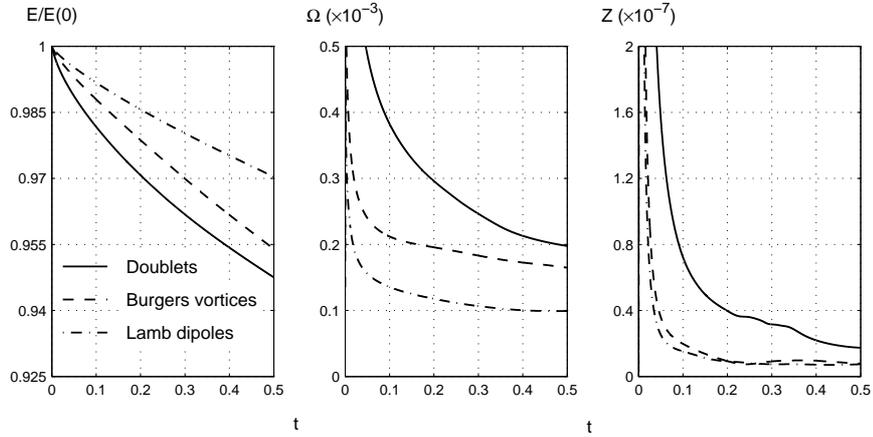}}
 \caption{Comparison of initial data (\ref{doublet}), (\ref{burgers}), and (\ref{lamb}) at $\nu{=}10^{-4}$. In practice, we could have synthesised these data in such a way that, say, the three flows start with an equal characteristic velocity or they have a fixed length scale. Evidently, a single static dimensionless number based on these parameters clearly does not differentiate the three core structures and the specific eddy orientations. In experiments, it is probably easy to generate the doublets  configurations but specialised test techniques may be required to instigate the other two types. } \label{cmphist} 
\end{figure}

Although the square's walls distort, to some extent, the path and appearance of the vortices, the broad features of leapfrogging and the subsequent merging of the vortices of identical signs are generic and highly representative. Suppose that we are attempting to determine a set of possible similarity solutions; we would like to be able to describe the start of leapfrogging and the instance of complete vortex merging, for instance. The ``obvious'' linear dimensions would be the vortices' separations in the $x$ or $y$ direction, or the initial core sizes, while the ``obvious'' velocity scale is the initial starting velocity of either vortex pair, given the fact that we do not have the {\it a priori} knowledge of a mean velocity at any instant $t>0$. Clearly, difficulties arise here as how to choose a characteristic length or velocity. We do not know whether there is a definitive Reynolds number of particular relevance for our purposes, see also figure~\ref{cmphist}. 

\begin{figure}[ht] \centering
  {\includegraphics[keepaspectratio,height=11.5cm,width=11.5cm]{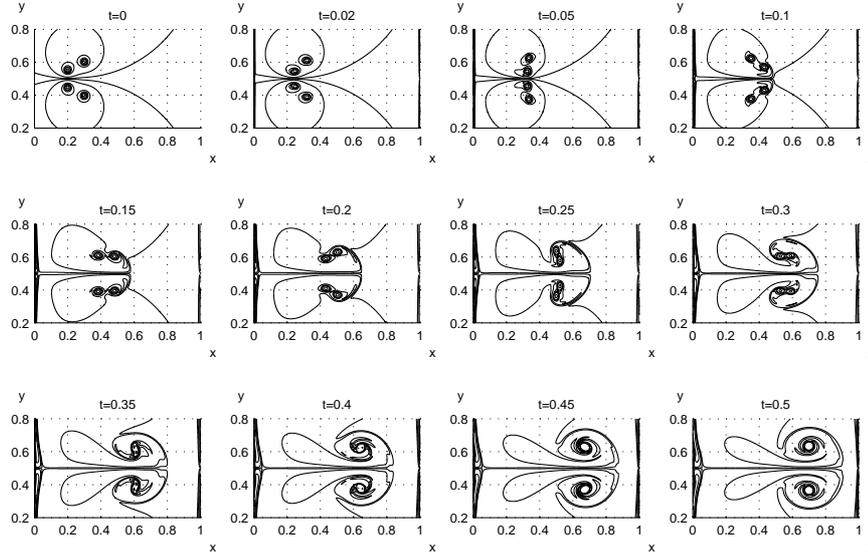}}
 \caption{The left doublets with a smaller separation given by expression (\ref{dbtsm}) 
 (cf. figure~\ref{dbteq10k}, $\nu=10^{-4}$). The initial separation clearly determines the time at which the overtake and the subsequent fusion occur. } \label{dbtsm10k} 
\end{figure}

\begin{figure}[ht] \centering
  {\includegraphics[keepaspectratio,height=11.5cm,width=11.5cm]{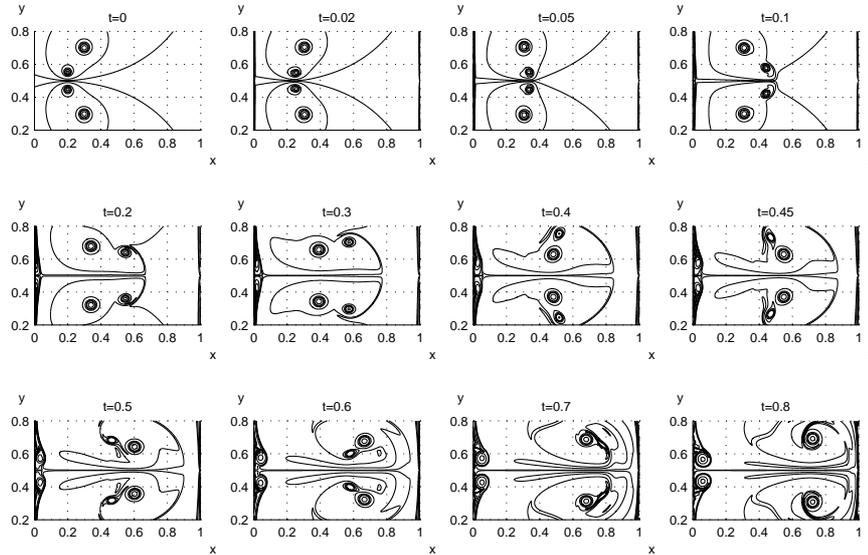}}
 \caption{The right doublets with a larger separation (\ref{dbtlx}) compared to the configuration of figure \ref{dbteq10k}. The fast-moving inner pair exert strong induction on the slow-moving ones, so that the initial large separation is reduced, for instance, at $t>0.1$ as the inner pair shoot to the right. Evidently, if the separation is large enough and the initial spin of the vortices is weak, leapfrogging may not occur at all and the pairs simply advect to the right until they are dissipated by viscous forces. } \label{dbtlx10k} 
\end{figure}

Our computational results establish that when two pairs are getting close to each other, the swirl due to their mutual spinning interaction expedites strongly. The weaker vortices are then smeared, elongated and absorbed by the stronger ones to a new counter rotating pair which continues to travel to the right wall. For fixed viscosity, the rate of the annihilation process must depend on the relative strengths of the eddies. Specifically, the present calculations, for instance, figure~\ref{dbtsm10k} or figure~\ref{dbtlx10k}, should be compared with the flow visualisation presented in Plate $79$ of van Dyke (1982). Every detail of the merging process can be clearly identified while in the experiment, the mixing core (interpreted in terms of the streak lines) appeared to have been blurred by the dye. 

Helmoholtz's insight asserts that every fluid motion, characterised by its vorticity, is an absolutely {\it unsteady} phenomenon of given initial conditions, regardless whether viscosity is included or not. Any initial solenoidal data of finite energy are perfect candidates for the vortex rings. We must pay attention to the fact that, in broad general circumstances, vortex leapfrogging or agglutination is neither purely laminar nor solely turbulent, as confirmed by experiments. In reality, thanks to the unlimited scope of start-ups, the consequential non-linearity captures the multifariousness of flow spectacle encountered in Nature and laboratory.   
\vspace{0.25cm}
\begin{acknowledgements}
\noindent 
23 July 2018

\noindent 
\texttt{f.lam11@yahoo.com}
\end{acknowledgements}
%
%
%%%%%%%%%%%%
%\vspace{2cm}
%%%%%%%%%%%%
%
%plots
%
%
%Doublets:
\newpage
\newpage
\begin{figure}[ht] \centering
  {\includegraphics[keepaspectratio,height=12cm,width=12cm]{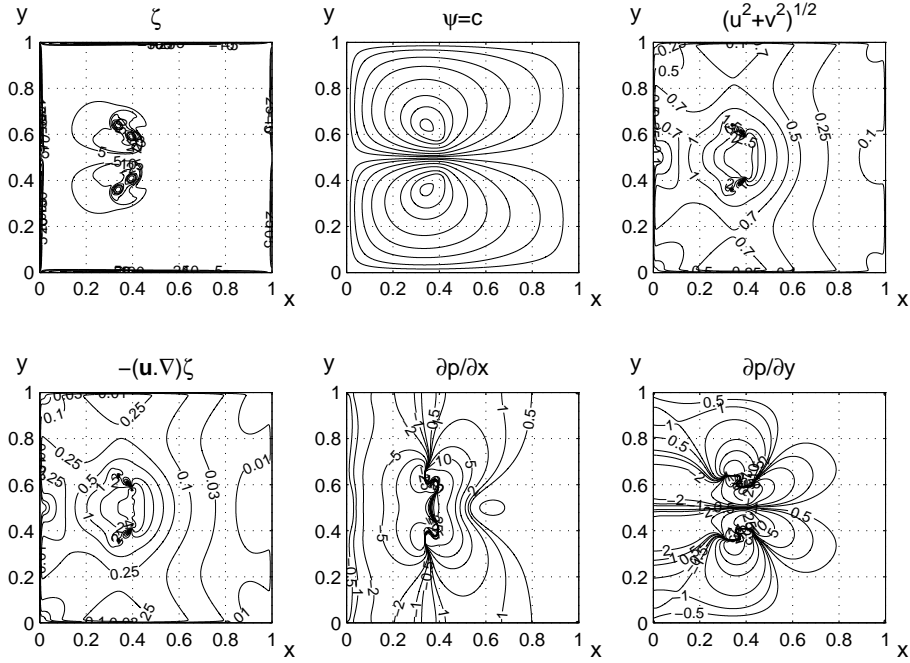}}
 \caption{Solutions at $t=0.12$ (doublets in tandem of figure~\ref{dbteq10k}), assuming unit density. The wall viscous layers are indeed thin and highly localised. Note that the velocity at the centres is subject to the mutual self-induction of the vortex pairs. Due to the rotation of the vortices, the pressure gradients have an S-shaped core manifold.} \label{dbteq10ks} 
\end{figure}
\begin{figure}[ht] \centering
  {\includegraphics[keepaspectratio,height=6cm,width=12cm]{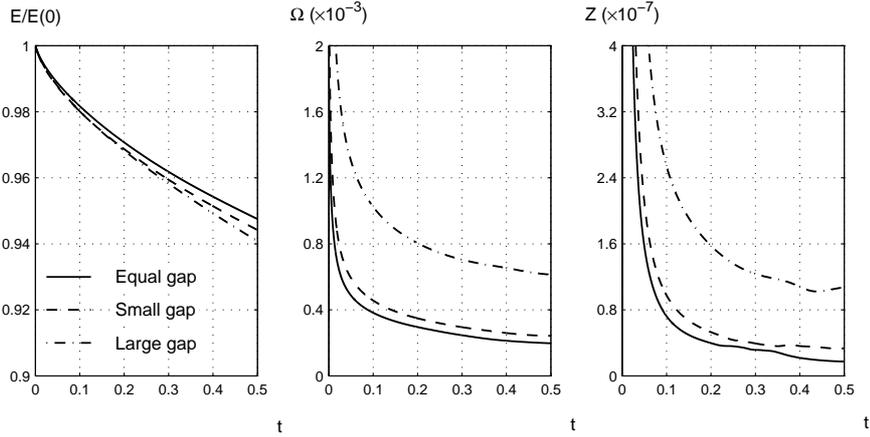}}
 \caption{Doublets at three gap arrangements, figures \ref{dbteq10k}, \ref{dbtsm10k} and \ref{dbtlx10k} (fixed $\nu=10^{-4}$). The quantities, $E/E(0)$, $\Omega$ and $Z$, are energy (normalised), enstrophy and palinstrophy respectively.} \label{dbthist} 
\end{figure}
%
%
%Burgers:
\newpage
%
%
%Lamb:
\newpage
%
%
%Identical rings head-on collision:
\newpage

%
%
% Identical doublets head-on collision:
\newpage
%
%
%Doublet in merging:
%
%
%\clearpage
%Doublets peeling-off:
\newpage
\end{document}